\documentclass[10pt,twocolumn,final]{IEEEtran}
\usepackage[dvips,final]{graphicx}
\usepackage{latexsym}
\usepackage{amssymb}
\usepackage[cmex10]{amsmath}
\usepackage{amsthm}
\usepackage{bm}
\usepackage{bbm}
\usepackage{array}
\usepackage{balance}
\usepackage[caption=false,font=footnotesize]{subfig}
\usepackage{cite}
\usepackage{algorithmic}
\usepackage{algorithm}
\usepackage{multirow}
\usepackage{pgfplots} 
\usepackage{color}
\usepackage{enumitem}
\usepackage{subfiles}
\usepackage{commath}
\usepackage{nicefrac}
\usepackage{wrapfig}
\usepackage{multirow}
\usepackage{tabu}
\usepackage{subfig}

\allowdisplaybreaks

\pgfplotsset{compat=newest}
\pgfplotsset{plot coordinates/math parser=false}

\title{Viewport-Aware Deep Reinforcement Learning Approach for 360$^o$ Video Caching}

\author{Pantelis Maniotis,
	    Nikolaos~Thomos, \IEEEmembership{Senior~Member,~IEEE}%
\thanks{P. Maniotis and N. Thomos are with the School of Computer Science and Electronic Engineering, University of Essex, Colchester, United Kingdom (e-mail: \{p.maniotis, nthomos\}@essex.ac.uk).}
}

\begin{document}
\maketitle

\begin{abstract}
360$^o$ video is an essential component of VR/AR/MR systems that provides immersive experience to the users. However, 360$^o$ video is associated with high bandwidth requirements. The required bandwidth can be reduced by exploiting the fact that users are interested in viewing only a part of the video scene and that users request viewports that overlap with each other. Motivated by the findings of our recent works where the benefits of caching video tiles at edge servers instead of caching entire 360$^o$ videos were shown, in this paper, we introduce the concept of virtual viewports that have the same number of tiles with the original viewports. The tiles forming these viewports are the most popular ones for each video and are determined by the users' requests. Then, we propose a proactive caching scheme that assumes unknown videos' and viewports' popularity. Our scheme determines which videos to cache as well as which is the optimal virtual viewport per video. Virtual viewports permit to lower the dimensionality of the cache optimization problem. To solve the problem, we first formulate the content placement of 360$^o$ videos in edge cache networks as a Markov Decision Process (MDP), and then we determine the optimal caching placement using the Deep Q-Network (DQN) algorithm. The proposed solution aims at maximizing the overall quality of the 360$^o$ videos delivered to the end-users by caching the most popular 360$^o$ videos at base quality along with a virtual viewport in high quality. We extensively evaluate the performance of the proposed system and compare it with that of known systems such as Least Frequently Used (LFU), Least Recently Used (LRU), First-In-First-Out (FIFO), over both synthetic and real 360$^o$ video traces. The results reveal the large benefits coming from proactive caching of virtual viewports instead of the original ones in terms of the overall quality of the rendered viewports, the cache hit ratio, and the servicing cost.
\end{abstract}


\section{Introduction}
\label{sec:intro}
Interactivity in VR/AR/MR systems is facilitated by the use of 360$^o$ video content. However, the interactivity associated with 360$^o$ videos comes with a huge increase in the bandwidth needed to deliver the content to the users. This puts pressure on the network infrastructure demanding further investments to accommodate 360$^o$ video related network traffic. Exploiting 360$^o$ video coding flexibility, i.e., encoding in tiles, and caching at the edge servers, can be a remedy for the problem, as we have shown in \cite{ManiotisMMSP19,ManiotisTMM20}. However, existing solutions assume known popularity, which may not always be the case. Further, existing solutions do not scale well with big content because of the cache optimization complexity. This naturally calls for caching systems that exploit tiles encoding and can estimate future content popularity trends, while preserving low complexity and scalability with respect to the number of 360$^o$ video files.

In 360$^o$ videos, a 360$^o$ view of a scene is captured from a single point with the use of an omnidirectional camera. The captured scene is then mapped to the internal part of a spherical surface. Each user is assumed to be placed at the center of the sphere and is interested in watching only a portion of the scene, known as viewport. Typically, each viewport covers 120$^o$ of the entire scene. According to the head movements of the user, the Head Mounted Display (HMD) dynamically alters the part of the scene that will be displayed. To prevent users from experiencing motion sickness and discomfort, the response of the system to the head movements should be as fast as the HMD refresh rate \cite{Viewport-adaptive}. Considering that the refresh rate may be 120Hz, the whole system should project the requested viewport in less than 10ms. However, state-of-the-art network streaming architectures are not able to respond under these tight time constraints due to the end-to-end delay. Although transmitting the whole scene could help to overcome the above limitation, it is not an efficient strategy since the resolution of a 360$^o$ video is commonly 4K, 8K, or even higher \cite{360_Dataset}. Thus, it would lead to significant bandwidth waste as only a part of the $360^o$ video would be eventually displayed. 

In edge caching systems, Small Base Stations (SBSs), e.g., picocells and femtocells, are equipped with caches, which can store a limited amount of popular content files. This is inspired by the fact that only a small number of popular content accounts for most of the network traffic load \cite{Charact_of_YouTube}. As a result, when there are multiple content requests for a cached content at an SBS, these can be served from the cache directly instead of obtaining the content through the core network using pricey backhaul links. This allows users to receive the content with lower latency, and the use of the backhaul links is limited. The potential of using edge caching as a solution to address the challenges that 360$^o$ video delivery faces in cellular networks, has been recently studied in \cite{ManiotisMMSP19,ManiotisTMM20,Papaioannou}. These works showed that offline edge caching can be a prominent solution for 360$^o$ video delivery, in particular, when tiles and layered encoding are used. The main drawback of \cite{ManiotisMMSP19,ManiotisTMM20,Papaioannou} is that they assume that the content popularity profile is known in advance. However, often in practice, the content popularity changes dynamically and may not be known a priori, or the estimated distribution may not be accurate. For regular videos, this problem has been addressed by online caching schemes \cite{MAB_1, MAB_4}. These methods learn the optimal caching policy by observing previous video consumption patterns. Though these methods are efficient for standard videos, they cannot be applied straightforwardly for 360$^o$ video. This is because 360$^o$ videos have considerably larger sizes than traditional videos, which limits the number of videos that can be stored at the SBSs caches. Furthermore, online caching schemes for regular videos have not been designed to take advantage of the fact that large parts of the video scene are never displayed, as is the case of 360$^o$ video where users are interested in watching only a viewport. From the discussion above, it is clear that there is a vast need for online $360^o$ video caching schemes which exploit 360$^o$ video features and do not necessitate the delivery of the entire video scene.

\begin{figure}
    \begin{center}
    \subfloat[Encoding of $360^o$ video in two quality layers and several tiles. ]{{\includegraphics[width=9cm]{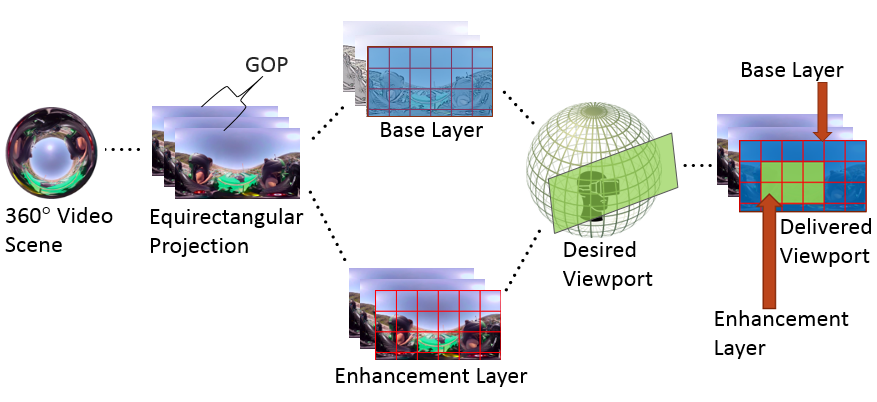} }\label{Encoding-a}}
    \qquad
    \subfloat[Overlapping viewports for various user requests, and virtual viewport.]{{\includegraphics[width=9cm]{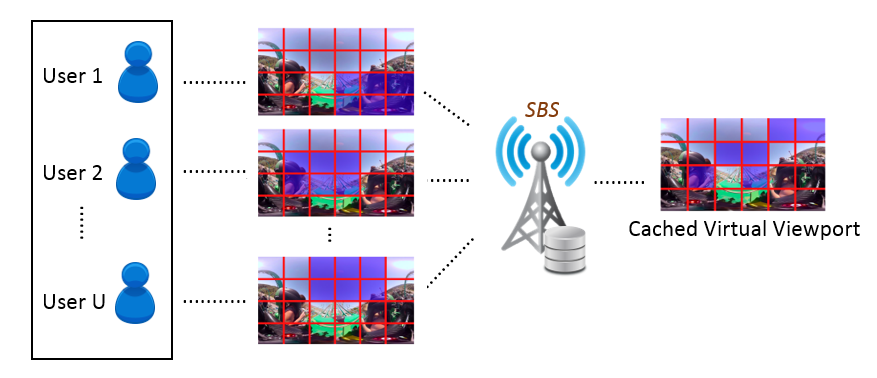} }\label{Encoding-b}}
    \end{center}
    \caption{Users request different viewports encoded in quality layers and tiles.}
\end{figure}

In this paper, we propose a proactive caching scheme for the transmission of 360$^o$ video in cellular networks. To the best of our knowledge, this is the first online caching scheme for $360^o$ videos. Our method aims at maximizing the overall quality of the 360$^o$ videos delivered to the users, without requiring \textit{a priori} knowledge of 360$^o$ video and tiles popularity distributions. To this aim, our method updates the cached content based on limited observations regarding 360$^o$ video consumption patterns, obtained from previous users' requests. We adopt tiles and layered encoding of 360$^o$ video because of the flexibility they offer to caching algorithms \cite{ManiotisMMSP19,ManiotisTMM20,Papaioannou}. These methods encode 360$^o$ videos into a number of independently encoded tiles and multiple layers, as shown in Fig. \ref{Encoding-a}. Encoding in tiles and layers allows network operators to cache in high quality at each SBS only the parts of the scene of each $360^o$ video (i.e., the tiles that correspond to these parts) that are the most popular to the users. Further, as only some tiles of the $360^o$ videos are popular, we introduce the concept of \textit{virtual} viewport, which is shaped by the overlap that occurs because of the diverse users' requests for different viewports, as shown in Fig. \ref{Encoding-b}. Virtual viewports differ from the original ones in that the tiles that comprise them are not necessarily adjacent to each other, i.e., they do not form a rectangular area. A virtual viewport has the same number of tiles with regular viewports, but it consists of the most popular ones. When a user requests a viewport of a $360^o$ video in a certain quality, then if some tiles of the requested viewport also belong to the virtual viewport that is cached at the SBS that received the request, these tiles will be served from that cache. As a result, storing virtual viewports will lead to an increase in the cache hit ratio, due to the greater flexibility they provide in terms of which tiles to cache in high quality. 

In order to determine which videos and virtual viewports to cache in each SBS, we first formulate the problem of $360^o$ video caching as a Markov Decision Process (MDP). The aim is to find the optimal set of $360^o$ videos and virtual viewports that should be cached at the SBS so that the overall quality delivered to the users is maximized. This is done by considering a limited history of users' requests. Although MDP offers an elegant way to describe our framework, the requirement of knowing the state transition probabilities makes it hard to evaluate the optimal policy (caching decisions per $360^o$ video and virtual viewport) for our system. This requirement can be lifted with the use of Q-learning \cite{Q_Learning}. Despite Q-learning convergence properties, it cannot be trivially applied for large-sized problems. To address this limitation of the Q-learning algorithm, we use the Deep-Q-Network (DQN) \cite{DQN-Arxiv} variant of Q-Learning. We evaluate the performance of our solution for both real \cite{Dataset} and synthetic $360^o$ video traces, and compare its performance with that of known schemes such as the Least Frequently Used (LFU), Least Recently Used (LRU), First-In-First-Out (FIFO) algorithms. The results illustrate the advantages of the proposed method compared to its counterparts, in terms of the overall quality users enjoy, the overall cache hit ratio, and the cost of delivering the requested content to the users.

In summary, the main contributions of our work are:
\begin{itemize}[leftmargin=*]

\item \textit{Reinforcement Learning framework:} We introduce a novel reinforcement learning framework for optimizing the content cache placement of
360$^o$ videos, by formulating the problem of caching 360$^o$ videos as a Markov Decision Process. Our solution aims at maximizing the overall video quality delivered to the users by taking into account both the $360^o$ videos and tiles' popularity.

\item \textit{Concept of Virtual Viewport:} We introduce the concept of the virtual viewport, which is shaped by the overlap of the diverse users' requests for different viewports. A virtual viewport is comprised of the most popular tiles of a $360^o$ video over the users' population. Virtual viewports enable us to reduce the size of the online cache optimization problem for $360^o$ videos.

\item \textit{Deep Neural Network representation:} We use Deep-Q-Network (DQN) to solve large instances of the online cache optimization problem for $360^o$ videos.

\item \textit{Evaluation on real and synthetic 360$^o$ video traces:} We extensively evaluate our proposed solution for real navigation patterns extracted from the dataset described in \cite{Dataset}, as well as on synthetic navigation patterns in order to show the benefits coming from the introduction of virtual viewports, and also, the impact of different users 360$^o$ video consumption patterns on the overall quality users enjoy.
\end{itemize}

The rest of the paper is organized as follows. In Section \ref{sec:rel_work}, we overview work related to edge caching, reinforcement learning, and tile-encoding of $360^o$ videos. Next, in Section \ref{sec:sys_setup}, we describe our system setup. Right after, in Section \ref{sec:modelling_user_requests} we introduce the considered model of the users' requests. Then, we first formulate our problem as an MDP in Section \ref{sec:MDP_formulation}, and right after in Section \ref{sec:Q-Learning_alg}, we show how DQN can be used to solve the cache placement problem for $360^o$ videos. In Section \ref{sec:sim_results}, we thoroughly evaluate the performance of the proposed scheme and compare it with other methods in the literature. Finally, we draw conclusions in Section \ref{sec:concl}.

\section{Related Work}
\label{sec:rel_work}
In this section, we briefly overview the literature related to edge caching, online caching, and tile-based $360^o$ video streaming.

The use of edge caching has been proposed as an efficient way to bring content closer to the end-users and improve the quality of the delivered content \cite{4k_Delivery, QoE-Edge_Caching_2}. In addition, caching popular contents at the mobile edge servers has been shown to reduce the usage of the pricey backhaul links \cite{Edge_Caching_1, Edge_Caching_2, Edge_Caching_3} and the network operation cost \cite{Edge_Caching_Cost}. The optimal placement of layered videos on edge caching systems is investigated in \cite{Rel_Work_Edge_Caching_1}. The decisions regarding which video layers to cache in each SBS are made by taking into account the caching cost, the available cache capacity at the SBSs, and the social groups formed by mobile users based on their content requests. Differently from \cite{Rel_Work_Edge_Caching_1}, caching several representations of multiple videos that correspond to different qualities is examined in \cite{QoE-Edge_Caching_2}. The cached representations are decided so that the aggregate distortion reduction of all the users is maximized while minimizing the cost related to downloading the representations. In \cite{4k_Delivery}, the delivery of 4K video quality in LTE-A networks is explored. That work aims to assure for 4K live streaming systems high Quality of Experience (QoE) to the users. 

The aforementioned works consider that video popularity profiles are known, which in many cases is not possible. To address this limitation, the content popularity is predicted using reinforcement learning algorithms that exploit the demand history \cite{MAB_1,MAB_4,ML1, ML2}. Specifically, in \cite{MAB_1}, the SBSs learn the content popularity online, considering the switching cost related to the addition of new files to the cache. 
Contextual MABs are proposed for online cache optimization in \cite{MAB_4} to take advantage of users' characteristics such as age, sex, etc. Neural networks (NN) \cite{ML1,ML2} can be used to decide the optimal cache placement when content popularity is unknown. Specifically, a Deep Reinforcement Learning-based framework aiming to maximize the long-term cache hit ratio is presented in \cite{ML1}. To limit the action space in \cite{ML1}, an Actor-Critic algorithm based on the Wolpertinger architecture \cite{ML0} is used. Differently, in \cite{ML2}, an Actor-Critic algorithm is presented where the actor uses the Gibbs distribution, and the critic uses a deep neural network to minimize the average transmission delay. To this aim, the users' scheduling and content caching policies are jointly designed. 

The delivery of $360^o$ videos encoded by advanced video coding standards, e.g., H.265/HEVC, SHVC, that support the encoding of the $360^o$ videos into a number of quality layers and tiles has been studied in \cite{Tile_approach_1,Tile_approach_2,Tile_approach_3}. These systems exploit the fact that users are interested in viewing only a viewport of the $360^o$ video scene, and hence there is no need to deliver the whole scene in high quality. Differently from \cite{Tile_approach_1,Tile_approach_2,Tile_approach_3}, in our previous work in \cite{ManiotisMMSP19,ManiotisTMM20} we proposed a tile-based collaborative caching scheme for $360^o$ videos for video-on-demand systems, where we showed the benefits coming from making the caching decisions on a per tile basis and the advantages of exploiting SBSs collaboration. In contrast to \cite{ManiotisMMSP19,ManiotisTMM20}, authors in \cite{Papaioannou} examine a tile-based caching scheme that aims to optimize the error between the requested and cached tile resolutions across different viewports as well as the coverage of the tiles set. In their work, they examine the caching of tile streams both at different resolutions and in a layered encoding fashion. The works in \cite{ManiotisMMSP19,ManiotisTMM20,Papaioannou} assume known popularity and hence cannot be used in a straightforward way for the problem studied here. A motion-prediction-based mechanism is proposed in \cite{Motion_Prediction}, where viewers' motion is predicted with the use of machine learning. Similarly, the navigation behaviors of users when watching $360^o$ videos on computers has been investigated in \cite{Study_360_Behaviours}. The results show that viewers have similar viewing patterns for certain $360^o$ video categories. A navigation-aware adaptive streaming strategy is presented in \cite{Navigation-Aware_Adaptive_Streaming}, where the aim is to optimize the rate at which a tile is downloaded during the navigation of the $360^o$ video. The rate per tile optimization problem is formulated as an integer linear programming problem. The proposed solution reveals the benefits of exploiting navigation patterns on both quality and navigation-smoothness. The impact of tile encoding on bandwidth saving, coding efficiency, and scalability is examined in \cite{Perf_measurements}, where a tile-aware video streaming system is proposed. The results show that an up to 80\% bit-rate reduction is achieved by only streaming the tiles viewed by the user.

\section{System Setup}
\label{sec:sys_setup}

In this section, we first introduce the system model and the network architecture, and then we discuss 360$^o$ video encoding into multiple quality layers and tiles. Finally, we present the employed viewport prediction algorithm.

\label{sec:sys_model}
\subsubsection{Wireless cellular network}
\label{sec:net_model}
\begin{figure}[t]
    \centering
	\includegraphics[width = 0.5 \textwidth ]{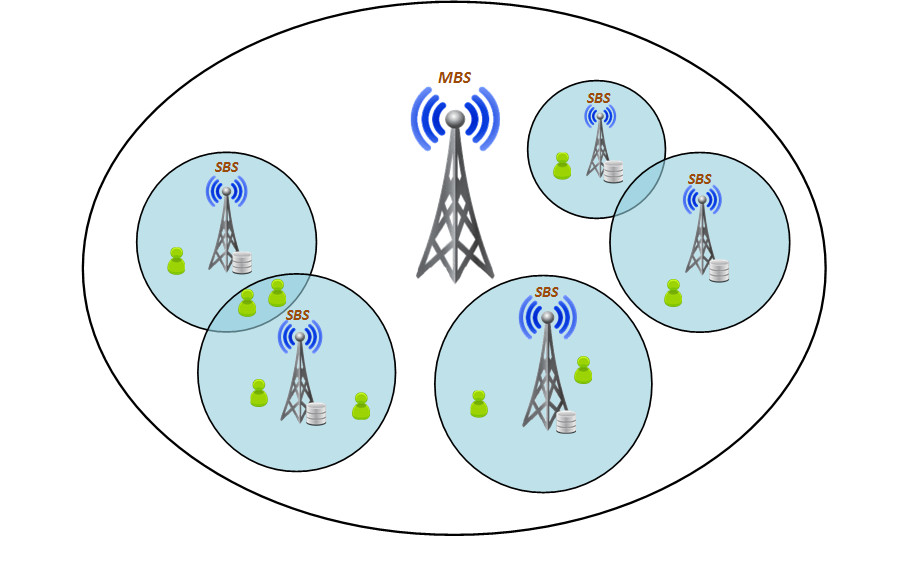}
	\caption{Considered network architecture.}
	\label{fig:Sys_setup}
\end{figure}

In this paper, we consider a heterogeneous cellular network (HCN), like the one depicted in Fig. \ref{fig:Sys_setup}. The network consists of $N$ Small Base Stations (SBSs), i.e., microcells, and a Macro-cell Base Station (MBS). Let $\mathcal{N}=\{1,\dots\,n,\dots,N\}$ denote the set of the $N$ SBSs, and $N+1$ represent the MBS. For notational convenience, we also define the augmented set $\mathcal{N}_B = \mathcal{N} \cup N+1$ that includes the SBSs along with the MBS. The MBS is connected to the core network through a high capacity backhaul link, i.e., optical fiber, while the SBSs are connected to the MBS through wireless millimeter-wave links.

Let $p_n$ be the communication range of the $n$th SBS and $\mathcal{P}=\{p_1,\dots\,p_n,\dots,p_N\}$ be the set that contains the communication ranges of all SBS. The communication range of the MBS is $p_{N+1}$, and is assumed to be large enough so that the MBS can communicate with all SBSs. Each SBS $n\in \mathcal{N}$ has a cache capacity $C_n \geq 0,\;\forall n \in \mathcal{N}$ where popular content can be cached. We further assume that there are $U$ users forming the set $\mathcal{U}=\{1,\dots\,u,\dots,U\}$. Since some users may be located in the overlap of the coverage areas of multiple SBSs, these users are assigned to the SBS with the maximum signal-to-interference-plus-noise ratio (SINR).

\subsubsection{Video Library}
\label{sec:video_library}
We assume that users request 360$^o$ video files from a content catalogue of $V=|\mathcal{V}|$ files, with $\mathcal{V}=\{1,\dots\,v,\dots,V\}$ being the set of the 360$^o$ videos. Each 360$^o$ video is encoded into $G$ Group of Pictures (GOPs) that form the set $\mathcal{G}=\{1,\dots,g,\dots,G\}$. Each GOP is encoded into $L$ quality layers and $M$ tiles. For each tile, the first quality layer is known as the base layer, while the rest $L-1$ layers are called enhancement layers. The acquisition of the base layer of a tile offers reconstruction of that tile at the lowest available quality, while the acquisition of all the layers of a tile up to the $l$th gradually improves the reconstruction quality of that tile. For each GOP, in order to satisfy a user demand for a requested viewport at a certain quality, the user has to acquire the base layer for all the tiles of the video along with all the enhancement layers corresponding to the demanded quality for the tiles that form the requested viewport.

\subsubsection{Viewport Prediction}
\label{sec:viewport_prediction}
A critical component of $360^o$ video streaming is the Viewport Prediction (VP) \cite{Flare, Two-Tier, OptTile}. The aim of VP is to predict the requested viewport by a user in the \textit{near} future (e.g., 1-2 sec), and prefetch it to the user. This is essential to provide smooth playback, as SBSs are not able to respond instantly to the user head movements due to the end-to-end delay.

VP can be done by observing the most recently requested frames by a user. These past requests are used to forecast the viewport that will be requested in the next few seconds. Such an approach is examined in \cite{Flare,Two-Tier}, where authors use variants of the linear regression algorithm to predict the users' head movements. A more na\"{\i}ve approach is presented in \cite{OptTile}, where VP is performed assuming that the users' head orientation will not change in the next 3 seconds. 

In our system, to perform viewport prediction, we use the Last Sample Replication (LSR) algorithm \cite{Two-Tier}. We have selected this algorithm because of its low complexity. Based on the LSR, the predicted viewport of the GOP $g+1$ is assumed to be the same as the one that was requested in the GOP $g$. For the first GOP, without loss of generality, we assume that the predicted viewport is the requested viewport. Although the employment of advanced VP algorithms \cite{Saliency1,Saliency2} would further improve the accuracy of the predicted viewports, we do not adopt such algorithms as we aim to show the advantages coming from caching. Further, the employment of more advanced prediction algorithms would increase the complexity of our system. At the same time, the conclusions derived regarding the benefits of tile-encoding and caching for $360^o$ videos would stay unaltered.

\subsubsection{End-to-end-delay} 
\label{sec:end-to-end} 
As we already mentioned, for each GOP $g \in \mathcal{G}$, all the tiles encoded at the base quality along with all the enhancement layers up to the targeted quality for the tiles that form the output viewport of the VP algorithm, need to be prefetched to the users within a specific time window. Failing to deliver these tiles on time would lead to buffer underruns, as the tiles would not be available to the buffer at the time they should be displayed. This would lead to degraded QoE, as tiles that are not delivered on time are discarded. Let us denote by $d_n$ the time needed to transmit one Mbit from the $n$th SBS to a user, and $d_{N+1}$ the time needed to transmit one Mbit from the backhaul of the MBS to a user. Obviously, $d_{N+1}>d_n$, due to the additional time needed to initially fetch data from the backhaul of the MBS to the SBS. The timely delivery of the tiles of each GOP must respect the following equation: 

\begin{equation} 
\small
	\sum_{n \in \mathcal{N}_B} \sum_{l \in \mathcal{L}} \sum_{m \in \mathcal{M}} o_{vglm} \cdot d_n \cdot q^{nu}_{vglm} \leq t_{disp}, \forall v \in \mathcal{V}, \forall u \in \mathcal{U}, g \in \mathcal{G}
	\label{endtoenddelay}
\end{equation}

\noindent where $o_{vglm}$ is the size of the $m$th tile encoded in the $l$th quality layer of the $g$th GOP of the $v$th $360^o$ video. The variable $q^{nu}_{vglm}$ takes the value $1$ when the $m$th encoded tile of the $l$th quality layer of the $g$th GOP of the $v$th $360^o$ video is delivered to the $u$th user from the cache of the $n$th SBS ($n \in \mathcal{N}$) or the MBS ($n=N+1$), and $0$ otherwise. The parameter $t_{disp}$ denotes the playback duration of each GOP. This constraint determines whether the tiles of the $(g+1)$th GOP can be prefetched as shown by the VP algorithm during the playback of the $g$th GOP. 

\section{Users' Requests Model and Cache Update Schedule}
\label{sec:modelling_user_requests}
In this section, we present the considered users' requests model and the cache update schedule. We assume that for each cached $360^o$ video, our system caches all the tiles at the base quality layer for all the GOPs, as well as the tiles of a $virtual$ viewport for each GOP in high quality. Recall that a viewport consists of $k$ tiles that form a rectangular area, while a virtual viewport is comprised of the $k$ most popular tiles, which do not necessarily form a rectangular area, as shown in Fig. \ref{Encoding-b}.

{\color{black}
We assume that time is slotted in $T$ time slots, and each time slot has the duration of one GOP. When a user is interested in watching a $360^o$ video with duration of $G$ GOPs, they should send $G$ consecutive requests,\footnote{When a user wants to stop watching a video, they halt sending requests for the following GOPs.} as shown in Fig. \ref{fig:Sys_Time}. The first request is special and comprises a request $w_0$, which is used by our algorithm in Section \ref{sec:Q-Learning_alg} to predict the popularity of each 360$^o$ video, and a request $w_{1}$ for obtaining the viewport for the first GOP. The request $w_0$ is to acquire the $360^o$ video at the base quality for all GOPs. As we will show in Section \ref{sec:Q-Learning_alg} this request is used to reduce the size of the optimization problem. Though all the tiles encoded at the base layer are requested at the first time slot, they may be delivered to the users along with the enhancement layer tiles. We assume that the request $w_0$ occurs at the same time slot with the request $w_1$. 
The $w_{g}$th request, $g \in \{1,\dots,G\}$, is to obtain the tiles that comprise the requested viewport, and belong to the $g$th GOP, in high quality. These requests are used by our algorithm presented in the next sections to calculate the popularity of each tile per GOP. For notational convenience, we denote the $i$th set of requests $\{w_0^i, w_1^i,\dots, w_G^i\}$ made by a user for a 360$^o$ video by $\mathcal{W}^i$, while $\mathcal{W}= \cup_{i} \mathcal{W}^i$ contains all the sets of users' requests. Hereafter, we drop the index of the $i$th set of requests when is not needed.

\begin{figure}[t]
    \centering
	\includegraphics[width = 0.55 \textwidth ]{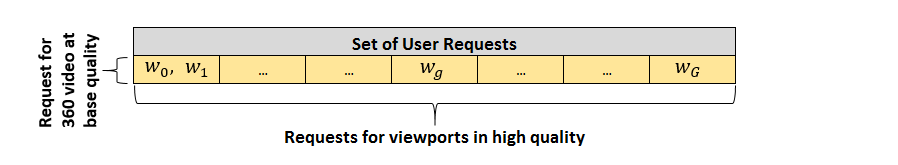}
	\caption{User requests for a $360^o$ video.}
	\label{fig:Sys_Time}
\end{figure}

The decisions of which tiles of a 360$^o$ video to cache at an SBS and in what quality are made online, i.e., when the content is requested. Specifically, when a user request $w_0$ arrives at an SBS in the time slot $1$, if the tiles of the requested 360$^o$ video at base quality are not cached in it, a decision has to be made regarding whether to cache them. This decision depends on the popularity of the video. If the decision is to cache these tiles, all the tiles of the base layer for all GOPs will start being fetched through the backhaul and cached at the SBS, replacing the tiles of another $360^o$ video that will be evicted. When the user issues further requests $w_g$ for receiving tiles of the GOPs of the 360$^o$ video in high quality, our system uses the viewport prediction algorithm described in Section \ref{sec:sys_setup} to decide which tiles of the predicted viewport will be fetched from the backhaul so that they can be delivered to the user on time. When the tiles that form the predicted viewport arrive at the SBS, we identify two cases: (a) if the decision for $w_0$ was not to cache the $360^o$ video in base quality, the fetched tiles will be delivered to the user but these tiles will not be cached at the SBS, as the video is not popular enough, (b) if the decision for $w_0$ was to cache the $360^o$ video in base quality, then for each request $w_g, g \in G$, in case some (or none) of the tiles that form the predicted viewport are cached in high quality a soft cache hit \cite{Soft_Cache_Hits} will occur. In such a case, the cached tiles of the predicted viewport will be served to the user directly from the SBS, while the tiles of the predicted viewport that are not cached at the SBS will be fetched to the SBS from a remote content server through the backhaul link of the MBS and be delivered to the user if the end-to-end constraint permits. Then, a decision is made about whether to cache some or all of the tiles that were fetched through the backhaul. The latter decision reflects tiles' popularity of a 360$^o$ video. 

}

The proposed cache optimization algorithm regarding which tiles to cache is presented in the next sections.

\section{MDP Formulation}
\label{sec:MDP_formulation}
In this section, we formulate the problem of caching $360^o$ videos in cellular networks as a Markov Decision Process \cite{MDP_0}. Since in our setting users can download the requested content only from the SBS that they are connected to, each SBS optimizes the cache use and the content replacement strategy independently of each other. Hereafter, following reinforcement learning terminology, SBSs are also called agents. 

\textit{State Space:} In the considered setting, the SBS $n\in \mathcal{N}$ can be in a state $s \in \mathcal{S}$, where $\mathcal{S}$ represents the set of all possible states. Each state is characterized by the features extracted from observations of users' past requests, considering fixed observation windows. Below we describe the features we consider.

The \textit{first feature} has two components that refer to the total number of requests for each cached 360$^o$ video that occurred in: a) a short-term window of $H_s$ sets of user requests (see Fig. \ref{fig:Sys_Time}), and b) a long-term window of $H_l$ sets of user requests. This feature associated with the cache of SBS $n\in \mathcal{N}$ can be described by the vector $\mathbf{x}^n=[\mathbf{x}^n_s\; \mathbf{x}^n_l]$ with $\mathbf{x}^n_f=[x^n_{f,i}]$, $\forall f\in \{s,l\}$, $\forall i\in{1,\dots,C}$, and $x^n_{f,i} \in \{1,\dots, H_f\}$. $x^{n}_{f,i}$ refers to the total number of times the video in the $i$th cache position was requested (either in short-term or long-term). Thus, the feature space $\mathcal{X}^n_f$ is given by $\{1,\dots,H_f\}^C$ and the overall feature space is $\mathcal{X}^n=\mathcal{X}^n_s \times \mathcal{X}^n_l$. Recall that, $C$ is the cache capacity of the SBS. It is worth noting that the above definition of features reduces the feature space drastically, as features are computed for all the tiles (cached videos) in base quality instead of each tile in base quality independently.

Similarly, the \textit{second feature} has two components that correspond to the total number of requests for tiles in high quality of the cached $360^o$ videos that happened during: a) the short-term window of $H_s$ sets of user requests, and b) the long-term window of $H_l$ sets of user requests. This feature is associated with the cache of SBS $n\in \mathcal{N}$ and is computed for each cached tile in high quality of GOP $g$, when request $w_g,\;g>0$ is processed. Let the vector $\mathbf{y}^n=[\mathbf{y}^n_s\; \mathbf{y}^n_l]$ describe this feature, where $\mathbf{y}_f^n=[y^n_{f,i,j}]$, $\forall f \in \{s,l\}$, $\forall j\in \{1,\dots,k\}$, $\forall i\in \{1,\dots, C\}$ and $y^n_{f,i,j} \in \{1,\dots, H_f\}$. $y^{n}_{f,i,j}$ denotes the number of times the $j$th tile of the $i$th cached 360$^o$ video was requested at the $n$th SBS. Thus, the feature space $\mathcal{Y}^n_f$ is given by $\{1,\dots,H_f\}^{kC}$ and the overall feature space for the cache space at the $n$th SBS is given by $\mathcal{Y}^n=\mathcal{Y}^n_s \times \mathcal{Y}^n_l$.

Finally, the \textit{third feature} has two components that correspond to the number of times the examined item (tile in high quality or  360$^o$ video in base quality) was requested at the $n$th SBS: a) in the short-term window of $H_s$ sets of user requests, and b) in the long-term window of $H_l$ sets of user requests. Specifically, when the examined item is a 360$^o$ video in base quality, this feature refers to the total number of times this video was requested at the $n$th SBS. This is the case when a request $w_0$ is received. When the  examined item is a tile of a 360$^o$ video in high quality, i.e. for requests $w_g,\; g>0$, the feature corresponds to the total number of times the examined tile was requested. 
The feature vector is defined as $\mathbf{z}^n=[\mathbf{z}^n_s\; \mathbf{z}^n_l]$ with $\mathbf{z}^n_f=[z^n_{f}]$, $\forall f\in \{s,l\}$, and $\forall z^n_{f} \in \{1,\dots, H_f\}$. $z^{n}_{f}$ stands for the total number of times the item (tile in high quality or  360$^o$ video in base quality) was requested. Thus, the feature space $\mathcal{Z}^n_f$ is given by $\{1,\dots,H_f\}$ and the overall feature space for the examined item is $\mathcal{Z}^n=\mathcal{Z}^n_s \times \mathcal{Z}^n_l$.

Following the above definitions of the features, the overall state space is given by:

\begin{equation} 
	\mathcal{S}^n= \mathcal{X}^n \times \mathcal{Y}^n \times \mathcal{Z}^n.
\end{equation}

Hereafter, we drop the superscript of the state space and use $\mathcal{S}$ as each SBS makes decisions independently of each other. 

\textit{Action Space:}
As we mentioned in Section \ref{sec:modelling_user_requests}, users' requests $w_0$ correspond to a request for a 360$^o$ video in base quality for all the GOPs of this video, while requests $w_g \in \mathcal{W}$ with $g \in {1,\dots,G}$ stand for a request for a viewport of the $g$th GOP encoded in high quality. 

When an SBS receives a request from a user, there are three possible cases regarding what data is cached at the SBS: a) no data for the requested $360^o$ video is cached, b) the $360^o$ video is cached at the base quality and the predicted viewport is cached at high quality and, c) the $360^o$ video is cached at the base quality, but a different viewport is cached at high quality.

In case a user requests a $360^o$ video that is not cached at the SBS, this has to be fetched through the backhaul and be delivered to the user. Fetching content through the backhaul adds cost to the network operator and increases the delay experienced by the users. When no data of a 360$^o$ video are cached at the SBS, a user request $w_g \in \mathcal{W}$ with $g \in \{0,\dots,G\}$ is processed as follows. To accommodate a request $w_0$, all the tiles of the requested $360^o$ video encoded at the base quality will start being fetched through the backhaul and delivered to the user for all the GOPs. For each of the following requests $w_{g}$ with $g \in\{1,\dots,G\}$, all the tiles of the viewport indicated by the prediction algorithm  in Section \ref{sec:sys_setup} encoded in high quality will be fetched through the backhaul in high quality, and be delivered to the user. Therefore, when the requested 360$^o$ video is not cached at the SBS, there are two types of possible actions: a) to leave the cached content at the SBS unchanged, or b) to evict the tiles of a cached 360$^o$ video from the cache of the SBS and replace them with the tiles of the requested one. Thus, there are $C+1$ possible actions. Let the set $\mathcal{A}_1=\{A_{10},A_{11}\dots,A_{1i},\dots,A_{1C}\}$ denotes all the possible actions when a video is not cached at the SBS. $A_{10}$ stands for the case the cached content at the SBS is left unchanged, and $A_{1i}$ means that all the tiles of the $i$th cached video at the SBS will be replaced by the corresponding tiles of the requested $360^o$ video. 

If both the requested 360$^o$ video encoded in the base quality and the tiles that form the predicted viewport for the examined GOP, e.g., $g \in \{1,\dots,G\}$ encoded in high quality are cached at the SBS, the request $w_g$ will be served from the cache, and no action will be taken. Then, a decision regarding whether to cache the tiles of the predicted viewport for the next GOP, i.e., $g+1$, is made. This happens because our scheme employs the LSR algorithm, as we described in Section \ref{sec:viewport_prediction}.

Finally, if the 360$^o$ video is cached at the base quality, but a different viewport than the predicted one is cached at the SBS at high quality, the requested tiles that are not cached have to be fetched through the backhaul, and then be served to the user. In that case, the possible actions are the following: a) to leave the cached viewport unchanged, or b) to cache some of the tiles, which were not part of the predicted viewport, and were fetched through the backhaul. To limit the action space, we assume that each action concerns only one tile that may be updated at the SBS cache. In this way, the agent takes sequentially actions for all tiles that were fetched through the backhaul in terms of whether to cache them at the SBS or not. This process is repeated until a decision is made for all the fetched tiles. Since each viewport consists of $k$ tiles, the possible actions for a tile form the set $\mathcal{A}_2=\{A_{20},A_{21}\dots,A_{2j},\dots,A_{2k}\}$. The action $A_{20}$ denotes the case where the cached content is left unchanged, while the action $A_{2j}$ corresponds to the case where the candidate tile will replace the $j$th tile in high quality of the requested $360^o$ video that was cached at the SBS. We consider that a GOP is fully processed when a decision has been made for all the tiles that were fetched through the backhaul. After completing the sequential decisions, the cached virtual viewport for the considered video is updated. Next, the subsequent GOP is processed in a similar way. We would like to note that the use of virtual viewports and the decomposition of actions on per tile basis permits to greatly reduce the action space as otherwise, the action space would have been comprised of all possible viewports.

Considering the above, the overall action space $\mathcal{A}$ is defined as:
\begin{equation} 
	\mathcal{A}= \mathcal{A}_1 \times \mathcal{A}_2.
\end{equation}

\textit{Reward:} 
We define the reward of each action to be the average distortion reduction the users will experience in the next $H$ sets of users' requests. Thus, given a state $s \in \mathcal{S}$, the reward of taking action $a \in \mathcal{A}$ is calculated as:
\begin{equation}
    \begin{aligned} 
    r(s,a)= \frac{1}{H}\sum_{ h \in {\mathcal{H}} } \sum_{ v \in {\mathcal{V}} } \sum_{g \in \mathcal{G}} \sum_{l \in \mathcal{L}} \sum_{m \in \mathcal{M}} \mathbbm{1} ( \phi_{h,v,g,l,m} ) \cdot \delta_{v,g,l,m}
    \label{eq:Reward}
    \end{aligned}
\end{equation}

When we process the $i$th set of requests $\mathcal{W}^{i}$, the set $\mathcal{H}=\{\mathcal{W}^{i+1},\dots,\mathcal{W}^{i+h},\dots, \mathcal{W}^{i+H}\}$ contains the next $H$ sets of user requests. In our formulation, the reward in (\ref{eq:Reward}) is obtained after the next $H$ sets of user requests have occurred \cite{sutton}. The term $\phi_{h,v,g,l,m}$ represents the $m$th tile of the $l$th quality layer of the $g$th GOP of the $v$th $360^o$ video of the $\mathcal{W}^{i+h}$th set of user requests. The term $\delta_{v,g,l,m}$ denotes the distortion reduction achieved by obtaining the corresponding tile. The indicator function $\mathbbm{1}(\phi_{h,v,g,l,m})$ in (\ref{eq:Reward}) is defined as:
\[
\mathbbm{1}(\phi_{h,v,g,l,m}) = 
\begin{cases} 
  1, & \text{if } \phi_{h,v,g,l,m} \text{ can be delivered}\\
      & \text{on time for $\mathcal{W}^{i+h}$.}\\
  0, & \text{if } \phi_{h,v,g,l,m} \text{ cannot be delivered}\\
      & \text{on time for $\mathcal{W}^{i+h}$.}
\end{cases}\nonumber
\]

\textit{Optimization Problem:} 

In order to quantify how good a particular state $s$ is, we estimate  the  value  function. This function corresponds to the expected discounted reward of policy $\pi$ when starting from a state $s$ and then following this policy. The value function is formally expressed as:
\begin{equation} 
	V_\pi(s)=E_\pi[G_\tau|S_\tau=s]=E_\pi[\sum_{\kappa=0}^\infty \gamma^\tau R_{\tau+\kappa+1} | S_\tau=s]
	\label{ValFunction}
\end{equation}
\noindent where $G_\tau$, $R_\tau$, and $S_{\tau}$ are the expected reward, the immediate reward and the state at time $\tau$, respectively. The parameter $0 \leq \gamma \leq 1$ is called discount rate and gradually discounts the effect of an action to future rewards. If $\gamma=0$, the agent is ``myopic" and maximizes the immediate reward. As $\gamma$ approaches 1, the objective takes into account future rewards more strongly, and the agent becomes farsighted. The above equation can be rewritten as a Bellman equation \cite{Bellman} as follows:
\begin{equation} 
	V_\pi(s)=\sum_a \pi(a | s) \sum_{s^\prime,r}p(s^\prime,r|s,a)[r+\gamma V_\pi (s^\prime)]
    \label{eq:Bellman}
\end{equation}
where $p(s^\prime,r|s,a)$ is the transition probability from the state $s$ to the state $s^\prime$ by taking the action $a$ with a reward $r$.

\section{DQN based cache optimization}
\label{sec:Q-Learning_alg}
The main challenge to solve (\ref{eq:Bellman}) is the requirement to know the transition probabilities $p(s^\prime,r|s,a)$. For the studied problem, continuous computation of the transition probability matrix is necessary because of the non-stationary requests' dynamics, which is computationally demanding. To overcome this problem, we can adopt the Q-learning algorithm \cite{Q_Learning}, which learns the optimal policy through interaction with the environment. Q-learning uses the $Q(s,a)$ values instead of using the value function in (\ref{eq:Bellman}). These values reflect how ``good" is to take action $a$ when in state $s$. Similarly, $Q_{\pi}(s,a)$ represents how good it is to take action $a$ when starting from state $s$, and thereafter follow the policy $\pi$. This is defined as follows:
\begin{equation} 
	Q_{\pi}(s,a)=E_\pi[\sum_{k=0}^\infty \gamma^\tau R_{\tau+k+1} | S_\tau=s, A_\tau=a]
\end{equation}
where $A_{\tau}$ is the action at time $\tau$.

The optimal policy is the one that maximizes the expected reward for all states and is given by:
\begin{equation} 
    \pi^\star(s)=\operatorname*{arg\,max}_{a \in \mathcal{A}} (Q(s,a)), s\in \mathcal{S}
\end{equation}

To determine the optimal policy $\pi^\star(s)$, the Q-learning algorithm updates the $Q(s,a)$ values iteratively. Specifically, the $Q(s,a)$ values are updated according to the formula:
\begin{equation} 
	Q(s_\tau,a_\tau) = (1-\alpha_{\tau})Q(s_\tau,a_\tau) + \alpha_{\tau}[R_\tau+\gamma \max_{a\in{\mathcal{A}}}	Q(s_{\tau+1},a)] 
	\label{q_target}
\end{equation}

\noindent where $\alpha_{\tau}$ is the learning rate at time $\tau$. The learning rate corresponds to the rate at which newly acquired information overrides old one.

Q-learning can select actions using policies such as the $\epsilon$-greedy, where $\epsilon \in [0,1]$, which ensures that random actions are always explored and overfitting is avoided. According to $\epsilon$-greedy policy, the action resulting in the maximum $Q(s_\tau,a_\tau)$ value is selected with probability $1 - \epsilon$, and a random action is selected with probability $\epsilon$. The Q-learning algorithm is guaranteed to converge to the optimal solution \cite{LuongTutorial2019} when all the state-action pairs are visited infinitely often, and the learning rate $\alpha_{\tau}$ satisfies the following conditions:
\begin{equation} 
\small
	\sum_{\tau=0}^\infty \alpha_\tau(s,a) = \infty \quad \mbox{and} \quad \sum_{\tau=0}^\infty \alpha_\tau^2(s,a) < \infty ,\quad \forall{(s,a) \in \mathcal{S} \times \mathcal{A}}
\end{equation}

The Q-learning algorithm is an efficient method to determine the optimal policy when the state-action space is small. However, when the state-action space grows, the lookup table where the $Q(s,a)$ values are stored becomes prohibitively large. To overcome this drawback of Q-learning, we employ a Deep Reinforcement Learning (DRL) \cite{DQN-Arxiv} approach. Using DLR the $Q(s,a)$ values are approximated by a Deep Neural Network (DNN). The DRL framework consists of two phases: a) the offline phase where the DNN is trained, and b) the online phase during which the actual caching decisions are made. 

During the offline phase, the DNN is initially built by selecting some random weights $\theta$. Then, the DNN is trained with a number of historic transition profiles, as in \cite{DQN-Measurements}. These profiles correspond to request patterns experienced in the past. The training of the DNN is performed in a mini-batch manner. Specifically, at each training epoch, a sample of the transition profiles and their estimated $Q$ values are obtained by randomly sampling the experience replay memory $D$, which has capacity $N_D$. This mechanism is used to remove the correlations between observations, while the transitions between the states become more independent and identically distributed.

To stabilize DNN training, apart from the experience replay, we use the mechanism of the fixed target network \cite{LuongTutorial2019}. According to this mechanism, a second DNN is employed, which is called fixed-target network. This network has the same architecture as the original DNN that is used for the function approximation (evaluation network). Not using a separate network to estimate the target $Q$ values would lead to destabilization. This would happen because as the $Q$ values (output of the evaluation network) are updated towards the target values (calculated by (\ref{q_target})), the target values will also be updated in the same direction. To overcome this problem, the weight parameters of the target network are kept fixed and are copied from the evaluation network only every $N_T$ steps. Thus, using a second network to estimate the target $Q$-values leads to a more stable training, since the $Q$-values obtained from the evaluation network are updated towards a target that is kept fixed (for a number of steps).

\begin{algorithm}[t]
\baselineskip=15pt
\caption{DRL Framework}
\label{alg:DQN}
\begin{algorithmic}[1]
\STATE \textbf{Offline Phase}
\STATE Initialize the evaluation network with weights $\theta$
\STATE Initialize the fixed target network with weights $\theta^{\prime}$
\STATE Initialize the experience buffer $D$ with capacity $N_D$
\STATE Initialize a random exploration process
\STATE Train the DNN with features $(s, a)$ and outcomes $Q(s,a)$ in a mini-batch manner
\STATE \textbf{Online Phase}
\FOR {each time slot}
\FOR {each user request in a time slot}
\FOR {each candidate item of a user request}
\STATE Receive observation $s_\tau$
\IF    {the candidate item is not cached at the SBS}
\STATE \textbf{With} probability $1-\epsilon$ \textbf{select} 
\STATE \hspace{5mm} $a_\tau=\arg \max\limits_{a\in\mathcal{A}} Q(s_\tau,a,\theta)$ 
\STATE \textbf{Otherwise,}
\STATE \hspace{5mm} $a_\tau$ $\gets$ random action
\STATE Take action $a_\tau$ and observe $r_\tau$, $s_{\tau+1}$
\STATE Store the tuple $(s_\tau,a_\tau,r_\tau,s_{\tau+1})$ in the experience replay buffer $D$
\ENDIF

\STATE Update cache hit ratio
\STATE Update Feature Space
\IF { Modulo($w$, $\mathcal{N}_B$)==0}
\STATE Sample $\mathcal{M}_B$ tuples from $D$ 
\STATE Update DNN by minimizing $Loss(\theta)$ in (\ref{Loss})
\STATE Update fixed target network weights
\ENDIF

\ENDFOR
\ENDFOR
\ENDFOR
\end{algorithmic}
\end{algorithm}

When the offline phase is completed, the obtained weights $\theta$ are used to initialize the DNN in the online phase. During this phase, if the candidate item ($360^o$ video in base quality or tile in high quality) is not cached at an SBS, the agent takes an action according to the $\epsilon$-greedy policy (i.e., it decides whether to cache the item or not and what content will be replaced), and then proceeds to the next state. In this way, new actions are always explored, and cached content whose popularity the algorithm overestimated in the past will not stay in the cache forever. After the execution of each action, the tuple $(s_\tau, a_\tau, r_\tau, s_{\tau+1})$ is stored in the experience replay buffer $D$, in order to be used later for the training of the DNN. 

In the online phase, the DNN is trained in a similar way to the offline phase, where a batch of $\mathcal{M}_B$ transition profiles is randomly sampled from the experience replay memory $D$ every $\mathcal{N_B}$ steps. The DNN is trained towards the target $Q$ values using the back-propagation method, by minimizing the loss function $Loss(\theta)$. The loss function is given by:
\begin{equation} 
	Loss(\theta)=\frac{1}{\mathcal{M}_B} \sum_{i\in\mathcal{M}_B} (y_i-Q(s_i,a_i,\theta))^2
	\label{Loss}
\end{equation}
where $y_i=r_i+{\max}_{a_i^{\prime}}Q(s_i^{\prime},a_i^{\prime},\theta_i^{-})$ represents the target Q value of the $i$th sample, and $\theta_i^{-}=\theta_{i-\mathcal{N_B}}$.

The overall DRL framework is presented in Algorithm \ref{alg:DQN}.

\section{Performance Evaluation}
\label{sec:sim_results}
In this section, we examine the performance of the proposed DQN-based online caching algorithm for $360^o$ videos in cellular networks. First, we describe the schemes under comparison and provide the simulation setup. Next, we show the convergence of the loss function during the training of the DNN. Then, we analyze the impact of various system parameters on the performance of the system. Finally, we demonstrate how the viewports' popularity shapes the popularity of each tile.

\subsection{Simulation Setup}
Let us describe the main characteristics of the schemes under comparison and the proposed scheme:

\begin{enumerate}
\item \textit{Least Frequently Used (LFU):} In this scheme, the network operator keeps track of the number of requests that occurred for each cached $360^o$ video. When a user request arrives at an SBS, then: a) if the requested $360^o$ video is not cached at it, all the tiles of the $360^o$ video that was requested the least number of times will be evicted from the cache of the SBS. Then, for all the GOPs, all the tiles of the requested $360^o$ video encoded at the base layer along with the tiles of the predicted viewport in high quality will be cached at the SBS; b) if the $360^o$ video is already cached at the base quality for all the GOPs, but some of the cached tiles in high quality are different from the ones that belong to the predicted viewport, these tiles will be evicted and be replaced by the tiles of the predicted viewport.

\item \textit{Least Recently Used (LRU):} In this scheme, the network operator keeps track of how recent are the requests that occurred for each cached $360^o$ video. When a user request happens at an SBS, then: a) if the requested $360^o$ video is not cached at the SBS, all the tiles of the $360^o$ video that were requested the least recently will be evicted from the SBS cache. Next, all the tiles of the requested $360^o$ video will be cached at the SBS at the base quality for all GOPs along with the tiles of the predicted viewport in high quality; b) if the $360^o$ video is cached at the SBS, for each GOP, if some of the cached tiles in high quality are different from the ones of the predicted viewport, these tiles will be replaced by the corresponding tiles of the predicted viewport.

\item \textit{First-In-First-Out (FIFO):} In this scheme, the network operator keeps track of when the requests for each cached $360^o$ video occurred. When a user request arrives at an SBS, then: a) if the requested $360^o$ video is not cached at the SBS, all the tiles of the $360^o$ video that was cached the earliest will be evicted from the SBS. Then, for all GOPs, all the tiles of the requested $360^o$ video encoded at the base layer, along with, for each GOP, the tiles of the predicted viewport in high quality will be cached at the SBS in the place of the evicted tiles; b) if the $360^o$ video is cached at the SBS, then for each GOP, if some of the cached tiles in high quality are different from the ones forming the predicted viewport, these tiles will be evicted, and be replaced by the tiles of the predicted viewport.

\item \textit{Proposed Scheme:} In the proposed scheme, the caching decisions are made exploiting observations derived from past users' requests. This scheme employs the DQN algorithm presented in Section \ref{sec:Q-Learning_alg} to decide on the cache updates. For each cached $360^o$ video, all the tiles at the base quality along with the most popular tiles in high quality that form a virtual viewport, are cached at the SBS for all the GOPs.

\end{enumerate}

For the sake of simplicity, all the conducted experiments are done assuming a single SBS and an MBS. This does not affect the derived conclusions, as SBSs make caching decisions independently of each other. As we have already mentioned in Section \ref{sec:sys_setup}, although SBSs' coverage area may overlap, users are assigned to a single SBS, i.e., the one with the maximum SINR. The exploitation of opportunities arising because of the overlapped coverage areas is part of our future work. We would like to emphasize that our algorithm can be applied to networks with an arbitrary number of SBSs. This is because as each user is assigned to a single SBS, our algorithm can run in parallel for each SBS. The coverage range of the SBS is set to be $P_n=300$m, while the coverage range of the MBS is $P_{N+1}=2000$m, and is large enough to permit the communication with the SBS. The delay needed to obtain one Mbit from the SBS is $d_n=1/6$ sec/Mbit, while the delay to deliver one Mbit from the backhaul of the MBS to the user is $d_{N+1}=1/2$ sec/Mbit. The cache capacity of the SBS is set to be enough to store 10\% of the $360^o$ videos of the content library. The number of users is $U=200$ who are randomly placed in the coverage area of the SBSs. Recall that, when a $360^o$ video is cached at the SBS, this means that for each GOP, all the tiles are cached at the base quality, and the tiles that form a virtual viewport are cached in high quality.

\begin{figure}[t]
	\centering
		\includegraphics[width = 0.5 \textwidth]{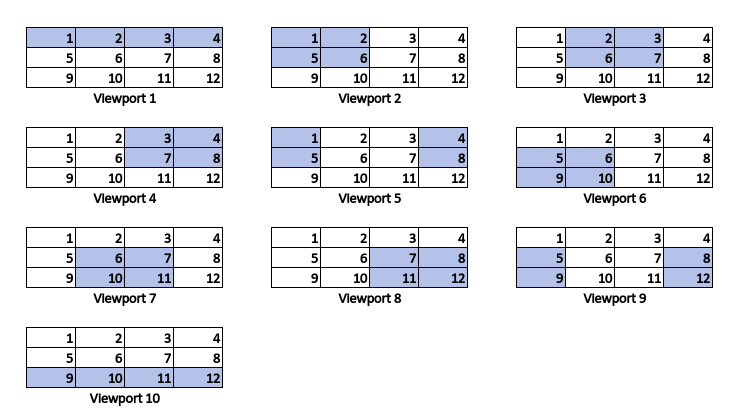}
		\caption{Considered set of viewports. The light blue area highlights the area covered by the viewports.}
	\label{fig:Viewports}
\end{figure}

The content library contains $V=500$ videos, while each video is encoded in 30 GOPs. The duration of each GOP is assumed to be $t_{disp}=1$ sec. Each GOP is encoded into $M=12$ tiles, where each tile is encoded into $L=2$ quality layers. The bitrate of the base layer is $2$ Mbps, while the bitrate of the enhancement layer is $12$ Mbps. The size of each viewport consists of $4$ tiles, while the available viewports are the ones depicted in Fig. \ref{fig:Viewports}. The distortion reduction achieved by obtaining a tile at the base quality layer is $30$ dB, while the distortion reduction achieved by receiving a tile at the enhancement quality layer is $10$ dB. The probability of a $360^o$ video to be requested from a user follows the Zipfian distribution \cite{zipf}, as it is common to the literature. The shape parameter of the Zipfian distribution is set to $\eta_v=1$. The probability of a $360^o$ video $v\in \mathcal{V}$ to be selected under the Zipfian distribution is given by: $$p_v=\frac{1/v^{\eta_v}}{ \sum_{v \in \mathcal{V} }1/ v^{\eta_v}}.$$ 

We consider realistic navigation patterns, extracted from the dataset in \cite{Dataset}, from which we sampled 200 trajectories of head movements. These trajectories are obtained from 10 different videos, where for each video, we sampled 20 different trajectories. With equal probability, we mapped the index of each one of the $V=500$ videos from the content library to one of the 10 sampled videos of the dataset. Then, for each of the $V=500$ videos of the content library, according to its mapped index, we selected one of the 20 available trajectories uniformly at random.

We assume that the total number of sets of users' requests is $W=10000$. The short-term time window refers to $H_s=300$ sets of user requests,\footnote{Each set of requests corresponds to the tiles of a single video demanded by a user.} while the long-term time window corresponds to $H_l=1000$ sets of user requests. The reward in (\ref{eq:Reward}) is calculated for the next $H=1000$ sets of user requests.

\subsection{Deep Neural Network Training}
\begin{figure}[t]
	\centering
		\includegraphics[width = 0.48 \textwidth]{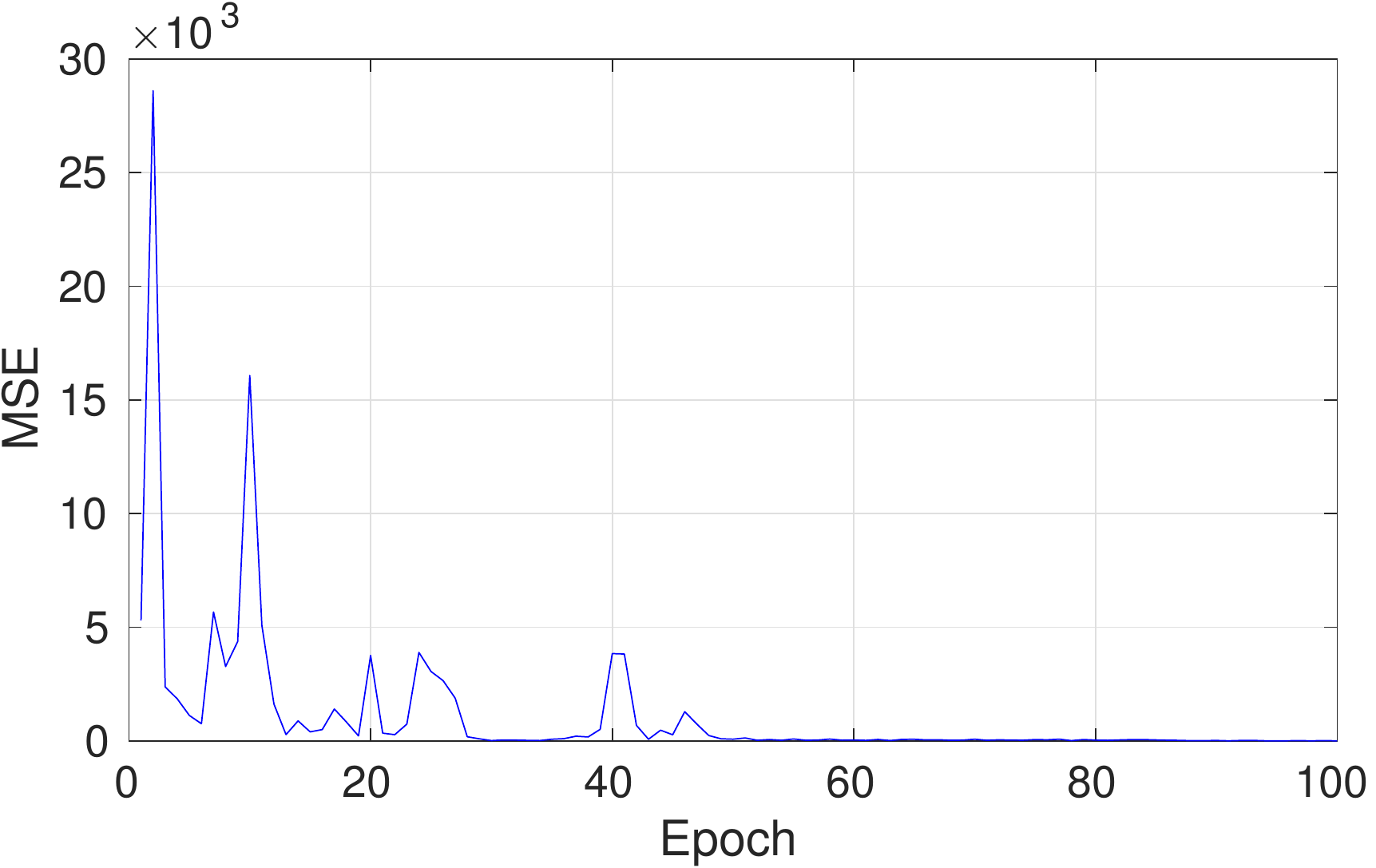}
		\caption{MSE of the loss function with respect to the training epochs.}
	\label{fig:Loss}
\end{figure}

We consider a Deep Neural Network (DNN), which consists of four fully connected layers, i.e., the input layer, two hidden layers, and the output layer. As the cache capacity of our system is $C$, the input layer consists of $10C+2$ nodes that reflect the vector size of each state. The hidden layers and the output layer consist of $5C+1$ nodes, as there are $5C+1$ total actions. The activation function of the hidden layers is the ``ReLu'', while the activation function of the output layer is the ``linear'' function. The DNN is trained with the \textit{Adam} optimizer. The DNN is trained for $100$ epochs in order to become sufficiently accurate. The learning rate is set to be $\alpha=0.001$, while the $\epsilon$-greedy parameter is set to $\epsilon=0.05$. The discount factor is set to be $\gamma=0.6$. The experience replay buffer is set to be $D=2000$, while the mini-batch size is set to $\mathcal{M}_B=32$. The mini-batch samples are obtained every $\mathcal{N_B}=200$ requests. The convergence of the loss function during the training phase for the basic scenario is presented in Fig. \ref{fig:Loss}. When the DNN is trained with different system settings than the ones of the basic scenario, a similar convergence behavior is noticed. 

\subsection{System Parameter Analysis}
\subsubsection{Cache Size} 
\begin{figure}[t]
	\centering
		\includegraphics[width = 0.48 \textwidth]{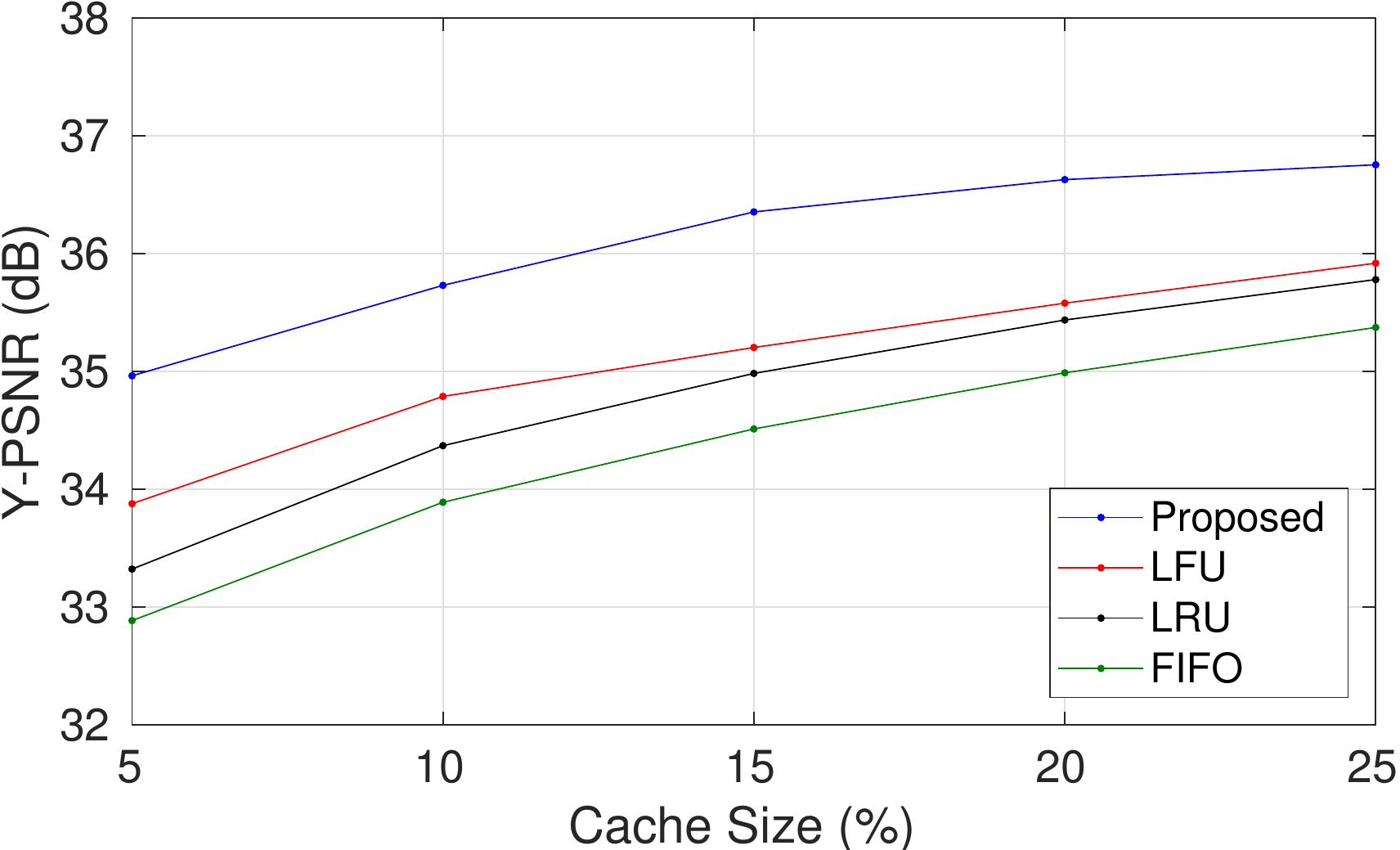}
		\caption{Y-PSNR of the rendered viewports with respect to the cache size for all the schemes under comparison.}
	\label{fig:Comp_CacheQ}
\end{figure}

\begin{figure}[t]
	\centering
		\includegraphics[width = 0.48 \textwidth]{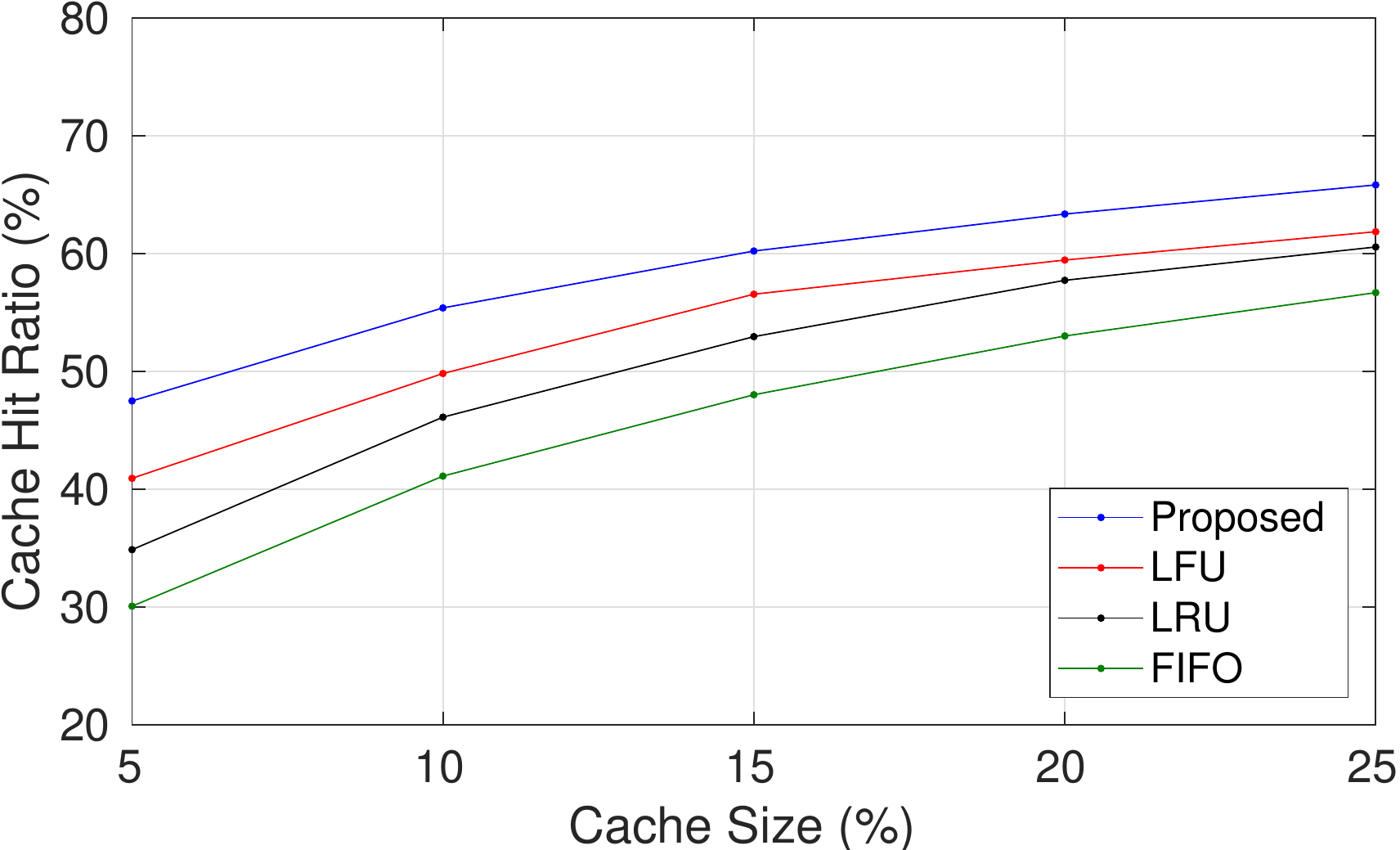}
		\caption{Cache Hit Ratio with respect to the cache size for all the schemes under comparison.}
	\label{fig:Comp_Cache}
\end{figure}

First, we examine the impact of the cache size on the overall quality of the rendered viewports. To this aim, we vary the cache capacity $C$ in the range [5, 25]\% of the size of the content library. As we can see in Fig. \ref{fig:Comp_CacheQ}, the proposed scheme outperforms the LFU, LRU and FIFO schemes, for all cache sizes. In particular, for typical cache capacity sizes, i.e., [5-10]\%, the performance gap between the proposed scheme and the LFU, the LRU and the FIFO is about $1$ dB, $1.5$ dB, and $2$ dB, respectively. This is because the proposed scheme achieves a better cache hit ratio, as shown in Fig. \ref{fig:Comp_Cache}. The increased cache hit ratio of the proposed scheme is attributed to the use of the DQN that learns from the experience of the past observations, which content should be cached. In addition, unlike LFU, LRU and FIFO, where the cached tiles in high quality correspond to actual viewports, in the proposed algorithm, the tiles that will be cached for each $360^o$ video in high quality correspond to virtual viewports. This provides us with greater flexibility to decide the cached tiles. The effect of the increased cache hit ratio on the quality of the rendered viewports comes from the fact that the tiles that are delivered from the cache of the SBS to the users are delivered with a smaller delay. Hence, more tiles are delivered in total to the users under the considered tight time constraints. When the cache capacity is large, i.e., 25\%, the performance gap between the proposed algorithm and the LFU, the LRU and the FIFO schemes closes to about $0.8$ dB, $1$ dB and $1.4$ dB, respectively. This happens because as the cache capacity becomes larger, most of the popular content is stored in the SBS cache for all the schemes.

\subsubsection{Video popularity distribution} 
\begin{figure}[t]
	\centering
		\includegraphics[width = 0.48 \textwidth]{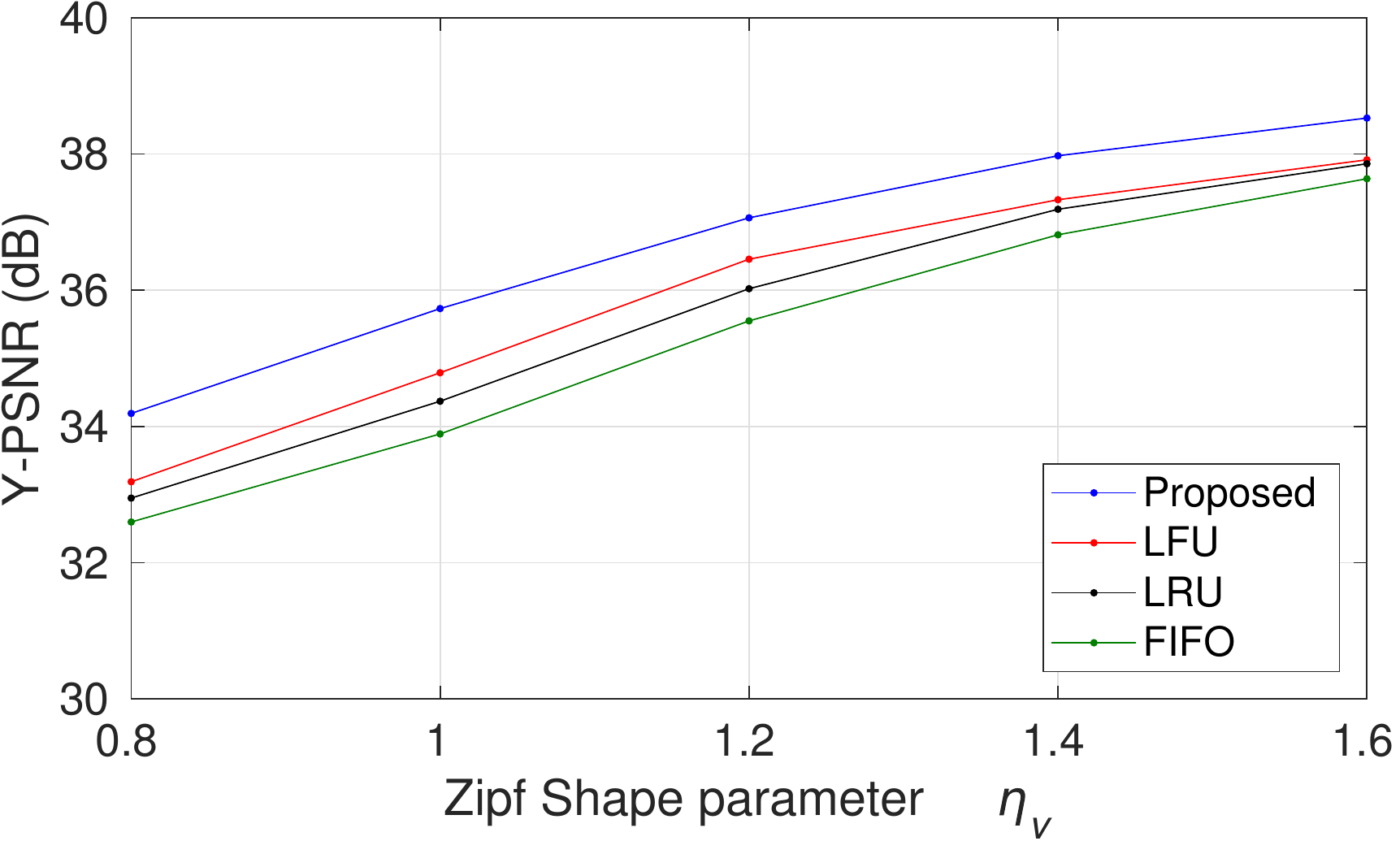}
		\caption{Y-PSNR of the rendered viewports with respect to the Zipf shape parameter of the $360^o$ videos for all schemes under comparison.}
	\label{fig:Comp_ZipfV}
\end{figure}

In Fig. \ref{fig:Comp_ZipfV}, we analyze the impact of the skewness parameter of the Zipfian distribution, which characterizes the $360^o$ video popularity. Specifically, we alter the shape parameter $\eta_v$ in the range [0.8, 1.6] and measure the overall quality of the rendered viewports for all the schemes under comparison. We note that an increase in the value of the Zipf shape parameter $\eta_v$ leads to an increase in the overall rendered quality for all the schemes. This is because bigger values of $\eta_v$ mean that the video popularity distribution gets steeper, i.e., a smaller number of $360^o$ videos is popular, which increases the efficiency of the cache utilization. We can further observe that as the users' requests concern a smaller number of videos (big $\eta_v$ values), the performance gap between the proposed algorithm and the LFU, the LRU, and the FIFO schemes decreases. For example, as the skewness parameter changes from $0.8$ to $1.6$, the performance gap between the proposed algorithm and the LFU decreases from $\sim 1$ dB to $\sim 0.6$ dB. This is attributed to the fact that as a smaller number of $360^o$ videos becomes popular, most of these videos will be cached at the SBS for all the schemes.

\subsubsection{Viewports' popularity distribution} 
\begin{figure}[t]
	\centering
		\includegraphics[width = 0.48 \textwidth]{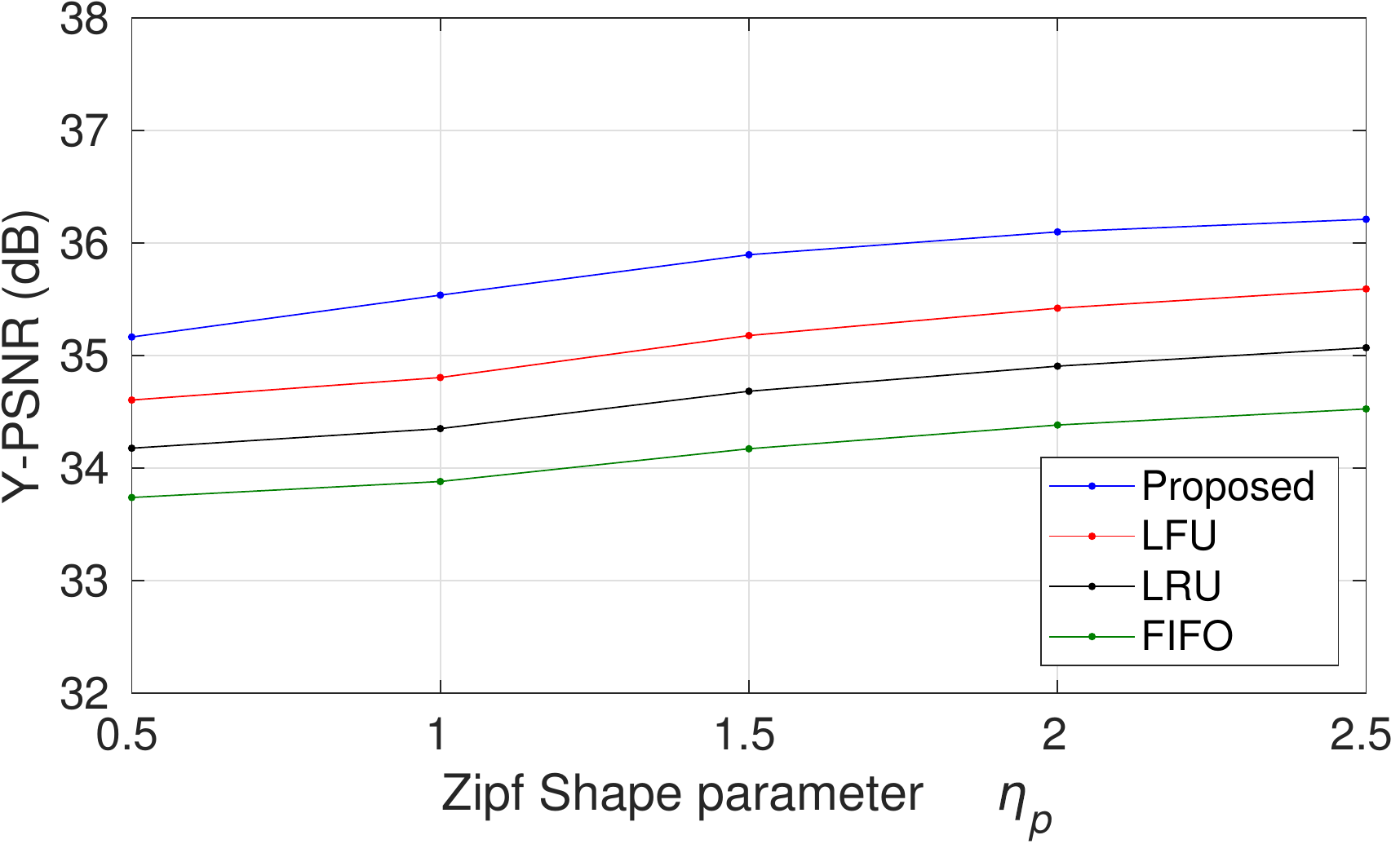}
		\caption{Y-PSNR of the rendered viewports with respect to the Zipf shape parameter of the viewports for all schemes under comparison.}
	\label{fig:Comp_ZipfT}
\end{figure}

Besides video popularity, we examine the impact of viewports' popularity. We first assume that the viewports' popularity follows a Zipfian distribution with skewness parameter $\eta_p$. To analyze the impact of the skewness parameter on the quality of the rendered viewports, we vary the shape parameter $\eta_p$ in the range [0.5, 2.5]. The performance of the schemes under comparison is depicted in Fig. \ref{fig:Comp_ZipfT}. From the results, we can note that an increase of the skewness parameter $\eta_p$ leads to an increase in the overall quality of the rendered viewports for all the examined schemes. This is because as the parameter $\eta_p$ increases, the user requests for the various parts of the $360^o$ video scenes become less diverse. Thus, the cache effectiveness is improved. In addition, as the skewness parameter changes from $0.5$ to $2.5$, the performance gap between the proposed algorithm and the LFU increases from about $0.5$ dB to about $0.65$ dB, respectively. This is because unlike LFU, in the proposed algorithm, the caching decisions for the tiles that will be cached in high quality are made for virtual viewports, which offers increased flexibility in the caching decisions regarding which tiles to cache. Thus, as the requests for the various viewports become less diverse, the performance gains in the proposed algorithm increase. Similar conclusions can be drawn by comparing the proposed scheme with the LRU and FIFO schemes. 

\begin{figure}[t]
	\centering
		\includegraphics[width = 0.48 \textwidth]{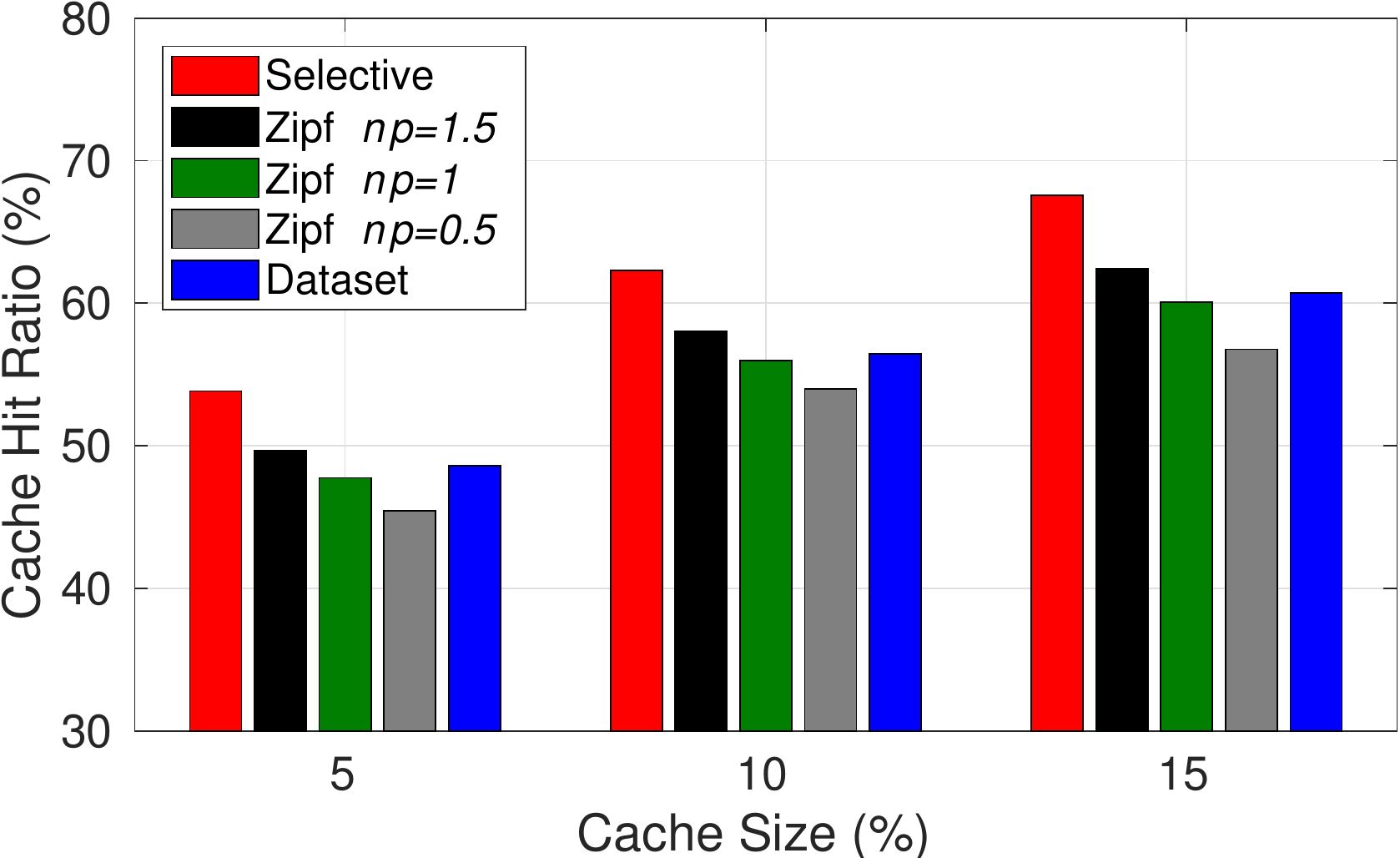}
		\caption{Cache hit ratio with respect to the cache size for the proposed scheme considering different viewport popularity distributions.}
	\label{fig:Comp_Viewp_Dist}
\end{figure}

In Fig. \ref{fig:Comp_Viewp_Dist}, we evaluate the cache hit ratio of the proposed scheme for: a) our basic scenario where the requests for the viewports are according to the dataset \cite{Dataset}, b) the case where the requests for the viewports follow the Zipfian distribution while the shape parameter $\eta_p$ takes a value from the range [0.5, 1.5], and c) the case where all the user requests are for one viewport, which we term as ``Selective". To this aim, we vary the cache size from 5\% to 15\% of the content library. As we can observe, the ``Selective" distribution achieves a better cache hit ratio in all cases. This is expected, as when the viewports follow either the dataset or the Zipfian distribution, the requests for the viewports are diverse, while in case of the Selective distribution, all requests are for one viewport. In addition, the cache hit ratio is better when the skewness parameter is higher as described above, while the performance of the dataset, is comparable with the case when the skewness parameter is $\eta_p=1$.

\subsubsection{Backhaul Usage} 
\begin{figure}[t]
	\centering
		\includegraphics[width = 0.48 \textwidth]{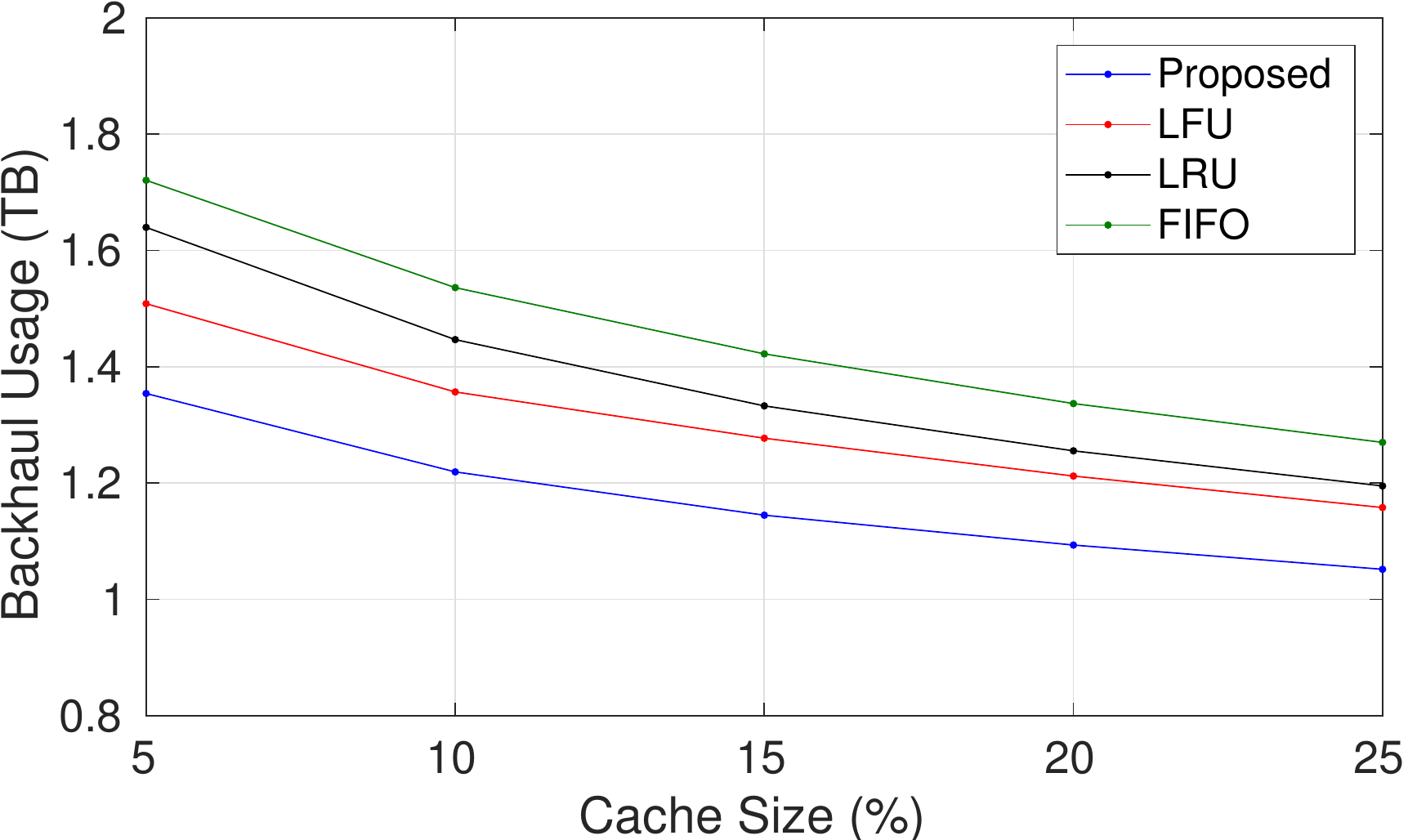}
		\caption{Backhaul usage with respect to the Cache Size for all schemes under comparison.}
	\label{fig:Comp_Backhaul_Usage}
\end{figure}

In Fig. \ref{fig:Comp_Backhaul_Usage}, we compare the performance of all the schemes under comparison in terms of the backhaul usage. This is a very important performance indicator of the caching schemes since field trials \cite{Subgradient_Poularakis} have shown that by reducing the backhaul usage, the network service cost is also reduced. To this end, we vary the cache size in the range [5, 25]\% of the content library and measure the backhaul usage, in terms of the bandwidth that should be communicated to satisfy the demands. As expected, an increase in the cache size leads to a decrease in the backhaul usage for all cases. This is because as the cache size increases, more videos will be able to be stored at the SBS cache, thus, more content will be served locally to the users. In addition, we can note that as the cache size increases, the performance gap between the proposed method and the other schemes under comparison decreases. Specifically, as the cache size increases from 5\% to 25\%, the performance gap between the proposed method and the LFU decreases from about $154.4$ GB to approximately $106.15$ GB. This is because as the cache size increases, most of the requested content will be able to be cached at the SBS, and thus, the effectiveness of the caching improves for all schemes.

\subsection{Overlap between Viewports}
\begin{figure}[t]
	\centering
		\includegraphics[width = 0.48 \textwidth]{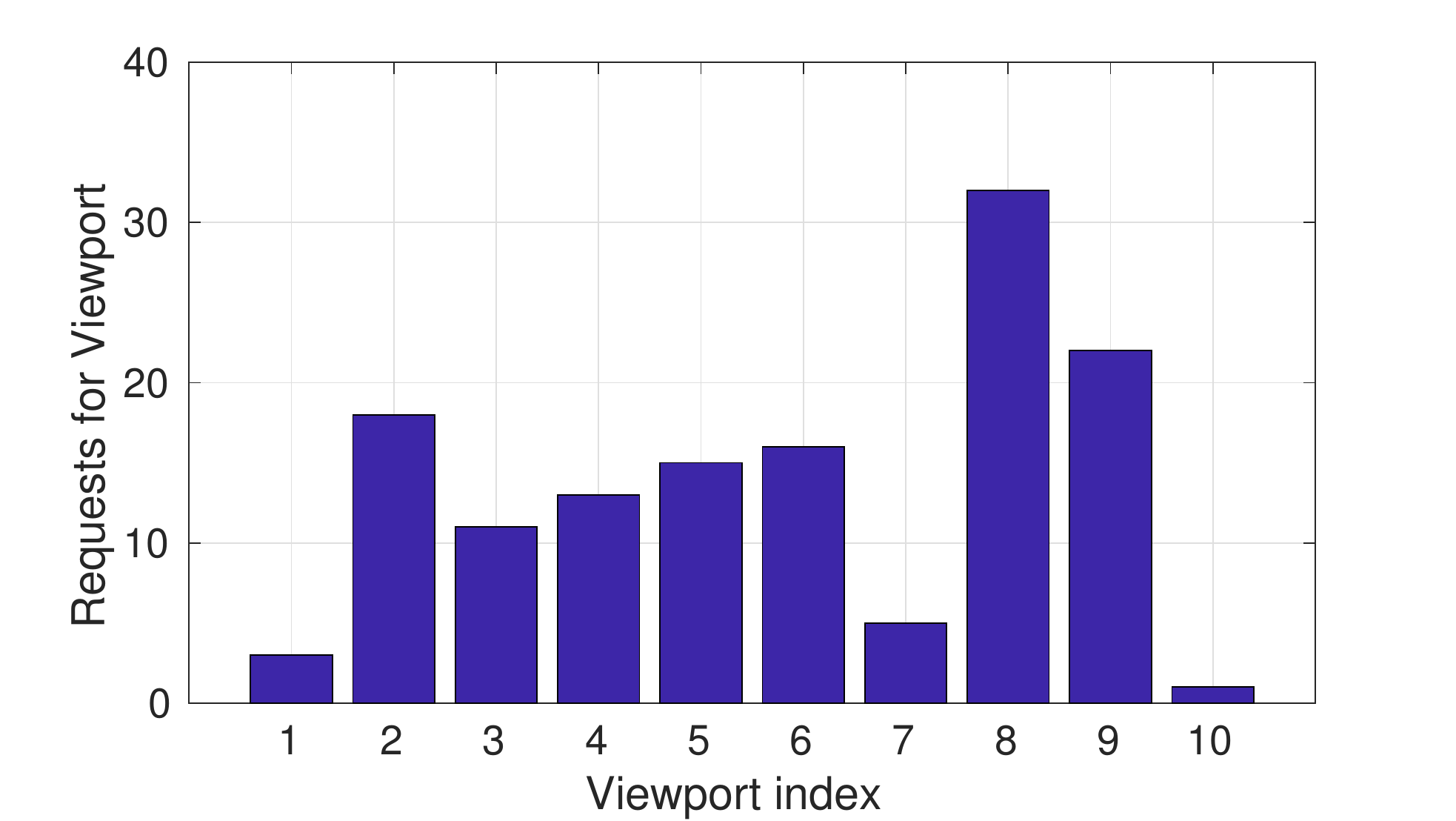}
		\caption{Total amount of requests for each one of the available viewports.}
	\label{fig:Comp_Viewp}
\end{figure}

\begin{figure}[t]
	\centering
		\includegraphics[width = 0.43 \textwidth]{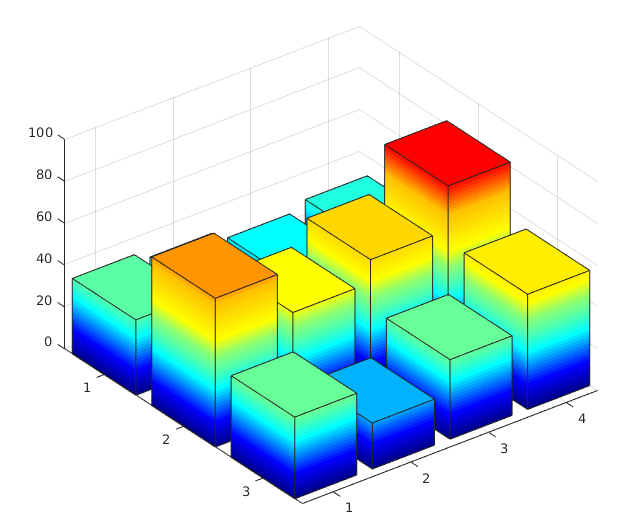}
		\caption{Total amount of requests for each one of the in high quality encoded tiles.}
	\label{fig:Comp_Tiles}
\end{figure}

In this section, we present how the overlap between the various viewports shapes the popularity of each tile. To this aim, we examine the popularity of each viewport, along with the popularity of each tile. 
These popularities are computed by measuring the frequency of occurrence of a request $w_g$ in a window of the previous $H_l=1000$ sets of user requests. The popularity of each viewport is depicted in Fig. \ref{fig:Comp_Viewp} and the popularity of each tile is depicted in Fig. \ref{fig:Comp_Tiles}. Although the most popular viewport is the viewport 8 (see the viewports illustrated in Fig. \ref{fig:Viewports}), by observing the Fig. \ref{fig:Comp_Tiles}, we can see that the most popular tiles do not correspond to the tiles of that viewport. The overlap between the diverse requests for the various viewports is what determines the popularity of each tile. Thus, by using \textit{virtual} viewports, which consist of the most popular tiles, the most popular tiles can be cached at the SBS. This results in higher cache hit ratio and better quality for the rendered viewports. 

\section{Conclusion}
\label{sec:concl}
In this work, we studied the problem of delivering $360^o$ videos in mobile networks using edge caching for unknown content popularity. We formulated the caching placement/eviction problem as a Markov Decision Process that aims at maximizing the overall quality of the videos delivered to the users. To deal with the dimensionality problem, we employ a DQN solution that exploits the patterns from the observations in the sequence of users' requests, in order to learn for each state, which cache update action should be taken. In this way, we are able to cache the $360^o$ videos that are predicted to be the most popular, along with for each GOP, a virtual viewport. To evaluate our method, we use both real and synthetic navigation patterns. We extensively compare our proposed method with the LFU, LRU, and FIFO schemes. The results show that the proposed method outperforms its counterparts. This improved performance is attributed to the exploitation of the tiles' popularity and the use of virtual viewports instead of the original ones, which increases the flexibility in the caching decisions.

\bibliographystyle{IEEEtran}

\end{document}